\documentclass[a4paper,10pt,twoside]{cpc-hepnp}

\usepackage{multicol}
\usepackage{graphicx}
\usepackage{booktabs}
\usepackage{amssymb,bm,mathrsfs,bbm,amscd}
\usepackage[tbtags]{amsmath}
\usepackage{lastpage}
\usepackage{CJK}

\usepackage{dcolumn}
\usepackage{float}
\usepackage{physics}

\begin{document}
\begin{CJK*}{GBK}{song}

\title{A systematical study of the chiral magnetic effects
	at the RHIC and LHC energies}

\author{%
	Bang-Xiang Chen$^{1}$
\quad Sheng-Qin Feng$^{1,2;1)}$\email{fengsq@ctgu.edu.cn}}

\maketitle

\address{%
	$^1$ College of Science, China Three Gorges University, Yichang 443002, China\\
	$^2$ Key Laboratory of Quark and Lepton Physics (MOE) and Institute of Particle Physics,\\
	Central China Normal University, Wuhan 430079, China}

\begin{abstract}
Considering the magnetic field response of the QGP medium, we perform a systematical study of the chiral magnetic effect(CME), and make a comparison it with the experimental results for the background-subtracted correlator $H$ at the energies of the RHIC Beam Energy Scan (BES) and LHC energy. The CME signals from our computations show a centrality trend and beam energy dependence that are qualitatively consistent with the experimental measurements of the charge dependent correlations. The time evolution of the chiral electromagnetic current at the RHIC and LHC energies is systematically studied. The dependence of the time-integrated current signal on the  beam energy $\sqrt{s}$ with different centralities is investigated. Our phenomenological analysis shows that the time-integrated electromagnetic current is maximal near the collision energy $\sqrt{s} \approx 39$ GeV. The qualitative trend of the induced electromagnetic current is in agreement with the CME experimental results at the RHIC and LHC energies.
\end{abstract}



\begin{multicols}{2}

\section{Introduction}\label{intro}

When two heavy ions collide with a nonzero impact parameter, a strong magnetic field with a magnitude of the order of $\mathrm{eB} \sim m_{\pi}^{2}$ \cite{Bloczynski:2012en,Skokov:2009qp,Voronyuk:2011jd,Bzdak:2011yy,Deng:2012pc,Mo:2013qya,Zhong:2014cda,Feng:2016srp}  ($ m_{\pi}$ is the pion mass), is generated in the direction of the angular momentum of the collision. The chirality imbalance should have experimental consequences in such a strong  magnetic field.
If the chirality is non-zaro, the quark spins are locked either parallel or anti-parallel to the magnetic field direction, depending on the quark charge.
 This would lead to the charge separation in the final state and to an electromagnetic current along the direction of the magnetic field
 \cite{Kharzeev:2007jp,Fukushima:2008xe,Muller:2010jd,Liu:2011ys}. Such charge separation and  electromagnetic current phenomena are called the chiral magnetic effect (CME) \cite{Liao:2010nv,Jiang:2014ura,Liao:2016diz,She:2017icp,Li:2018ufq,Kharzeev:2009pj}.

It has been argued that positive charges separate from negative charges along the direction of the angular momentum of the collision if the P and CP-violating processes occur in  QGP generated in relativistic heavy-ion collisions\cite{Bzdak:2019pkr,Kharzeev:2015znc}. The directional movement of positive and negative charges in a strong magnetic field should produce an electromagnetic current, which is an intriguing phenomenon that originates in the interplay of a quantum anomaly with the
magnetic field. The electromagnetic current $\overrightarrow{J} = \sigma \overrightarrow{B}$ would be induced by the chirality imbalance in an external magnetic field $\overrightarrow{B}$ , where $\sigma = e^{2}\mu_{5}/(2\pi^{2})$ is the chiral magnetic conductivity and $\mu_{5}$ is the chiral chemical potential.

Although there exists an obvious background contamination, it was suggested that signals of charge separation are seen in relativistic heavy ion collision data of
the STAR\cite{Abelev:2009ac,Abelev:2009ad,Adamczyk:2013hsi} and PHENIX \cite{Ajitanand:2010rc}experimental groups at RHIC and the ALICE \cite{Abelev:2012pa}collaboration at the LHC. With a new background subtraction method, the data obtained in the RHIC Beam Energy Scan (BES) \cite{Adamczyk:2014mzf} by the STAR experimental group further demonstrated the possible existence of the CME signal. It seems that the CME signal in the energy range from 19.6 to 62.4 GeV\cite{Adamczyk:2014mzf} is more clearer. A new phase of the RHIC energy scan  will be performed during 2020, which will allow a more accurate study of CME.

In this paper, we consider three important issues: 1) the magnetic field response of the quark gluon plasma (QGP) to the time evolution of the strong magnetic field; 2) the interplay of charge separation with the magnetic field; and 3) the dynamical processes in chiral magnetic current in response to the time-dependent magnetic field. We choose the simplified  KMW model to discuss the charge separation and to compare it with the experimental results at the RHIC and LHC energies. For the study of the electromagnetic current, we take into account the finite frequency response of CME to a time-varying magnetic field, find a significant impact of the QGP medium feedback, and study the generated electromagnetic current as a function of beam energy at the RHIC and LHC energies.

This paper is organized as follows: the magnetic field response of the QGP medium in relativistic heavy-ion collisions is given in sect. 2. The charge separations at the RHIC and LHC energies is discussed in sect.3. In sect.4, we use the Kubo formula to compute the electromagnetic current at the energies of the RHIC BES, the top RHIC energy, and at the LHC energy 2.76 TeV. The conclusions are summarized in sect.5.

\section{The magnetic field with the response of QGP medium}\label{eb}
One of the main issues of CME is the time evolution of the magnetic field in relativistic heavy-ion collisions. This issue has been investigated in many studies\cite{Skokov:2009qp,Voronyuk:2011jd,Bzdak:2011yy,Deng:2012pc,Feng:2016srp,Kharzeev:2007jp,Shi:2017cpu,Jiang:2016wve,Huang:2017tsq}, which found that enormous magnetic fields $\left(\mathrm{B} \sim 10^{15} \mathrm{T}\right)$  can be generated at the very beginning of the collisions. However, according to these studies, the intensity of the magnetic field rapidly decreases with time. The higher the collision energy, the faster is the magnetic field decrease is problematic. Recently, a limit of the magnetic field effect at late times was reported in \cite{Guo:2019joy,Guo:2019mgh,Muller:2018ibh} by studying the chiral vortex effect in relativistic heavy-ion collisions. Nevertheless, it was suggested in \cite{Feng:2016srp,She:2017icp} that the calculation of the magnetic field in vacuum is appropriate only for the early stage of collisions, and that the magnetic field response of the QGP medium should be considered after the formation of QGP.

Tuchin studied \cite{Tuchin:2010vs} the magnetic field properties in the QGP medium and suggested that due to the large electric conductivity, the magnetic field is partially 'frozen' during the entire plasma lifetime. The magnetic conductivity of the QGP medium was also quantitatively studied in \cite{Tuchin:2013ie,Zakharov:2014dia,McLerran:2013hla,Tuchin:2014iua,Tuchin:2015oka}. We also made a study of the space-time evolution of the magnetic field in QGP in~\cite{Feng:2016srp,She:2017icp}. The magnetic field at the center of QGP has only the $y$ component, and the magnitude of the magnetic field is given as
\begin{equation}\label{equ.01}
B_{y}\left(t \geq t_{0}, \mathbf{0}\right)=\frac{t_{0}}{t} e^{-\frac{c_{s}^{2}}{2 a_{x}^{2}}\left(t^{2}-t_{0}^{2}\right)} B_{y}^{0}(\mathbf{0}) .
\end{equation}
where $t_0$ is the formation time of partons, $B_{y}^{0}(\mathbf{0})$ is the magnetic field at t = $t_0$ and at the central point $(\vec{r}=0)$, $c_s$ is the speed of sound, and $a_{x}$ is the root-mean-square of the transverse entropy distribution. Here, we use $c_{s}^{2} \sim {1}/{3}$ and $a_{x} \sim 3$ . The formation time $t_0$ is given as \cite{Feng:2016srp,She:2017icp}

\begin{equation}\label{equ.02}
t_{0} \simeq 1 / Q_{s} ,
\end{equation}
where $Q_{s}$ is the saturation momentum, which is

\begin{equation}\label{equ.3}
Q_{S}^{2} \sim A^{1 / 3} x^{-\varpi}, \quad x=Q_{S} / \sqrt{s},
\end{equation}
where $A$ is the atomic number of the colliding nucleus, and $\varpi$ is a parameter between 0.25 and 0.3 ($\varpi=0.3$ in this paper). The saturation momentum for different nuclei and center-of-mass energies is

\begin{eqnarray}\label{equ.4}
Q_{s}^{2}(\sqrt{s}, b, A)  &=&
\left(\frac{A}{197}\right)^{\frac{2}{3(2+\varpi)}}\left(\frac{\sqrt{s}}{130}\right)^{\frac{2{\varpi}}{(2+\varpi)}}\nonumber\\[1mm]
&&\times Q_{s}^{2}(\sqrt{s}=130GeV, b, A=197)
\end{eqnarray}
where $Q_{s}^{2}(\sqrt{s}=130 \textrm{GeV}, b, A = 197)$ was given in Ref. \cite{Kharzeev:2000ph}. The results for $t_0$ and  $B_{y}^{0}(\mathbf{0})$ for two centralities at the RHIC BES and LHC collision energies are given in Table 1.

\begin{center}
	\tabcaption{\label{tab.1}Results for $t_0$ and $B_{y}^{0}(\mathbf{0})$ for two centralities for collision energy at the RHIC BES and LHC energies.}
	\footnotesize
	\begin{tabular}{lcccc}
	\toprule
	$\sqrt{s}$(GeV) & \multicolumn{2}{l}{centrality: $10 \%-30 \%$}     & \multicolumn{2}{l}{centrality: $30 \%-60 \%$} \\
		\cline{2-5}
		\multicolumn{1}{c}{}  & \multicolumn{1}{c}{$t_{0}(f m)$} & $e B_{y}^{0}\left(M e V^{2}\right)$ & $t_{0}(f m)$  & $e B_{y}^{0}\left(M e V^{2}\right)$ \\
		\hline
		11.5&	0.209&	4275.0&	0.260&	6214.4\\
		19.6&	0.195&	6407.1&	0.242&	8045.0\\
		27  &	0.187&	7616.7&	0.232&	8687.4\\
		39  &	0.178&	8569.5&	0.221&	8753.0\\
		62.4&	0.168& 	8481.6&	0.208&	7653.7\\
		200 &	0.144& 	3980.8&	0.179&	2766.2\\
		2760&	0.102& 	579.3 &	0.126&	156.1\\
		\bottomrule
	\end{tabular}
\end{center}

\end{multicols}
\begin{center}
\includegraphics[width=18cm]{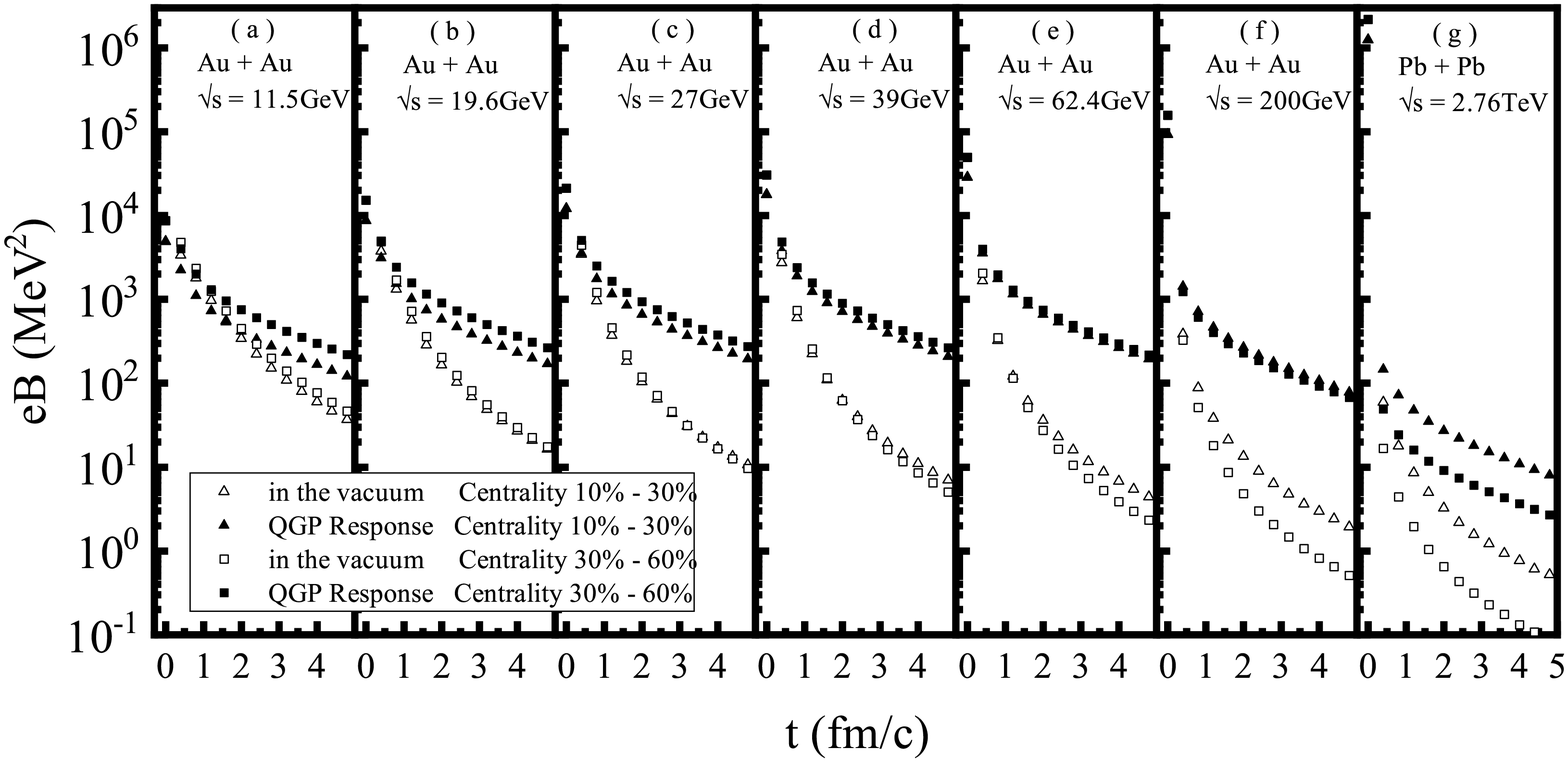}
\figcaption{\label{fig.1} Time evolutions of the magnetic field for two centralities in the Au - Au collisions, for $\sqrt{s}$ = 11.5. 19.6, 27, 39, 62.4, 200 GeV, and in the Pb-Pb collisions at $\sqrt{s}$  = 2760 GeV. The solid squares and solid triangles correspond to the results in the QGP medium and for centralities of $10 \%-30 \%$ and $30 \%-60 \%$,respectively. The hollow squares and hollow triangles are the results in vacuum with centralities  $10 \%-30 \%$ and $30 \%-60 \%$, respectively. }
\end{center}
\begin{multicols}{2}

The time evolutions of magnetic field is plotted in Fig.~\ref{fig.1} for two centralities at the RHIC BES energies, the top RHIC energy, and the LHC energy of 2.76 TeV.  The magnetic fields in vacuum at different energies are also plotted for comparison. Recently, the RHIC STAR collaboration~\cite{Adamczyk:2014mzf} presented the results of the dependence of charge correlations in the Au-Au collisions at midrapidity for center-of-mass energies of 7.7, 11.5, 19.6, 27, 39, and 62.4 GeV. It was observed \cite{Adamczyk:2014mzf} that the signal gradually reduces as beam energy is decreased, and tends to vanish below 7.7 GeV after background subtraction. This suggested that hadronic interactions dominate over partonic interactions at lower collision energies. Therefore, the chiral magnetic effect was analyzed starting from $\sqrt{s}=11.5$ GeV in the article.
It is found that the magnetic fields with QGP response last longer in the 27 - 62.4 GeV energy region. Compared with the magnetic field in vacuum, the lifetime of the magnetic field is longer when the QGP medium response is considered. The strength of the magnetic field decreases rapidly with time, and the higher the collision energy, the faster is the magnetic field decrease. Compared with the RHIC energies, the initial magnetic field (at $t = 0$) at the LHC energy is much bigger, but the magnetic field decreases much faster both in vacuum and with the QGP response. For the non-central collisions, the magnetic field is mainly due to the contribution of the spectator nucleus. When the two colliding nuclei are closer, the magnetic field generated is bigger, and for larger separations of the two nuclei, the magnetic field becomes smaller. For example, for the LHC energy, the spectator nucleus moves away almost at the speed of light, so that at higher collision energies, the magnetic field decreases faster.

\begin{center}
\includegraphics[width=9cm]{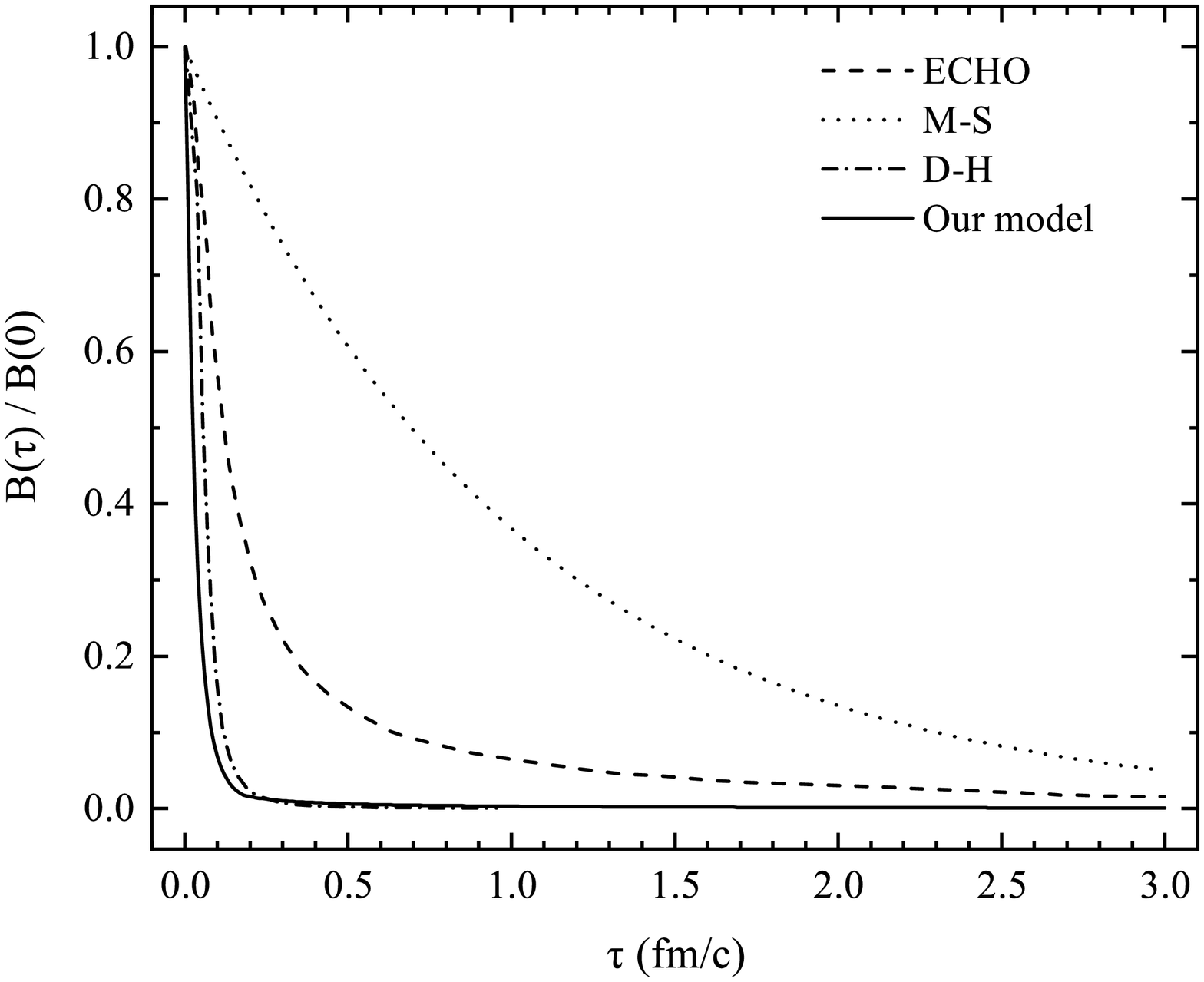}
\figcaption{\label{fig.2}
Comparison of the time evolution of the magnetic field, normalized to its peak value ,given in the studies: ECHO-QGP~\cite{Inghirami:2016iru} (dashed line curve), M-S~\cite{Muller:2018ibh} (dotted line curve), D-H~\cite{Deng:2012pc} (dash-dotted line curve), and in our model (real line curve).}
\end{center}

Figure 2 shows a comparison of the time evolution of the magnetic field normalized to its peak value, obtained in the studies by ECHO-QGP~\cite{Inghirami:2016iru}, M$\ddot{u}$ller and Sch$\ddot{a}$fer~\cite{Muller:2018ibh} (M-S) model, Deng and Huang~\cite{Deng:2012pc} (D-H) magnetic field calculation, and in our model. The magnetic field evolution in our model decreases more rapidily than the other models, and might induce a weaker CME signal. The results presented in the next sections are based on our model.

\section{Charge Separation at the RHIC and LHC Energies}\label{cme}
In this section, we first introduce the KMW model \cite{Kharzeev:2007jp}, and then give a detailed analysis of the CME in relativistic heavy-ion collisions at the RHIC and LHC energies.

The potential transition with non-zero winding number $Q_W$ passes through a barrier associated with QCD which exceeds the strong coupling constant $\alpha_{s}$. The transition can be implemented by an instanton \cite{Diakonov:2002fq,Schafer:1995pz} or sphaleron \cite{Arnold:1987zg,Fukugita:1990gb}. At low temperature, the transition is mainly achieved by the quantum tunneling effect, which is exponentially depressed by a transition called instanton. The transition at high temperatures is not forbidden and can be achieved by a transition called sphaleron. This may occur in the background of extremely high temperature quark gluon plasma (QGP). Thus, it provides a choice for generating chirality. On the other hand, the discovery of CME in relativistic heavy-ion collisions also implies generation of QGP.

The transition rate in QCD was given by the KMW model in Ref.~\cite{Kharzeev:2007jp} as follows:
\begin{equation}\label{equ.5}
\frac{\mathrm{d} N_{\mathrm{t}}^{ \pm}}{\mathrm{d}^{3} x \mathrm{d} t} \equiv \Gamma^{ \pm} \sim 192.8 \alpha_{S}^{5} T^{4},
\end{equation}
where the superscript $\pm$ defines the transition of $Q_W=\pm1$. The total transition rate is the sum of the rates of the ascending and descending transitions
\begin{equation}\label{equ.6}
\frac{\mathrm{d} N_{\mathrm{t}}}{\mathrm{d}^{3} x \mathrm{d} t}=\sum_{ \pm} \frac{\mathrm{d} N_{\mathrm{t}}^{ \pm}}{\mathrm{d}^{3} x \mathrm{d} t}.
\end{equation}

In the case of a suitable magnetic field with a large temperature $T$ and non-zero winding number $Q_W$, the charge separation given in Ref. \cite{Kharzeev:2007jp} is
\begin{equation}\label{equ.7}
Q \approx 2 Q_{\mathrm{w}} \sum_{f}\left|q_{f}\right| \gamma\left(2\left|q_{f} \Phi\right|\right)
\end{equation}
where
\begin{equation}\label{equ.8}
\gamma(x) = \begin{cases}
x, & \text{for } x \leq 1, \\
1, & \text{for } x \geq 1,
\end{cases}
\end{equation}
and $\varPhi=eB\rho^{2}$ is the magnitude of the magnetic flux.

We define by $N_{a}^{ \pm}$ and $N_{b}^{ \pm}$ the total positive/negative charge in units of $e$ above (a) and below (b) the reaction plane, respectively.  $\Delta_{ \pm}=N_{a}^{ \pm}-N_{b}^{ \pm}$   is the difference in charge between each side of the reaction plane. A charge difference will be generated locally when there is a transition from one vacuum to another. If the quarks experience many interactions in QGP, the observed final observed charge separation is suppressed. A suppression function
\begin{equation}\label{equ.9}
\xi_{ \pm}\left(x_{\perp}\right)=\exp \left(-\left|y_{ \pm}(x)-y\right| / \lambda\right)
\end{equation}
is introduced to describe nuclear screening, where $y_{\pm}(x)$ is the upper and lower $y$ coordinate of the overlap region, and $\lambda$ is the screening length. The expectation value of the change of  $\Delta_{+}$ and $\Delta_{-}$ due to a transition is either positive or negative with equal probability, and is given by
\begin{equation}\label{equ.10}
\pm \sum_{f}\left|q_{f}\right| \gamma\left(2\left|q_{f} \Phi\right|\right) \xi_{ \pm}\left(x_{\perp}\right),
\end{equation}
where only the most probable transitions $\left(Q_{W}=\pm 1\right)$ are considered.

One can calculate the variation of $\Delta_{ \pm}$ by assuming that all transitions occur independently from each other.
By using Eq.~(\ref{equ.5}) and $\rho \sim\left(\frac{\Gamma^{ \pm}}{\alpha_{s}}\right)^{-\frac{1}{4}} \sim 1 /\left(\alpha_{s} T\right)$, we calculate
$\langle \Delta^2_\pm \rangle$ and $\langle \Delta_+ \Delta_- \rangle$  for small magnetic fields $\left(2\left|q_{f} \mathrm{eB}\right| \leq 1 / \rho^{2}\right)$.
Since the magnetic field is a function of the rapidity $\eta$, one can compute $\left\langle\Delta_{ \pm}^{2}\right\rangle$ and $\left\langle\Delta_{+} \Delta_{-}\right\rangle$ as
\begin{align} \label{equ.11}
\langle \Delta^2_\pm \rangle &= 2\kappa \alpha_S \Bigl[ \sum_f q_f^2 \Bigr]^2 \int_{V_\perp} \dd[2]{x_\perp} \nonumber\\*
&\times [\xi_-(x_\perp)^2 + \xi_+(x_\perp)^2] \int_{\tau_i}^{\tau_f} \dd{\eta} \dd{\tau} \tau [eB(\tau, \eta, x_\perp)]^2,  \\ \label{equ.12}
\langle \Delta_+ \Delta_- \rangle &= -4\kappa \alpha_S \Bigl[ \sum_f q_f^2 \Bigr]^2 \int_{V_\perp} \dd[2]{x_\perp} \nonumber\\*
&\times \xi_+(x_\perp)\xi_-(x_\perp) \int_{\tau_i}^{\tau_f} \dd{\eta} \dd{\tau}  \tau [eB(\tau, \eta, x_\perp)]^2.
\end{align}
where the space-time rapidity is $\eta=\frac{1}{2} \log [(t+z) / t-z]$, and the proper time $\tau=\left(t^{2}-z^{2}\right)^{1 / 2}$. The magnetic field should not alter the transition rate dramatically. There is also a constant $\kappa$, of the order of magnitude of one  but with large uncertainties \cite{Kharzeev:2007jp}.
$\left\langle\Delta_{ \pm}^{2}\right\rangle$ and $\left\langle\Delta_{+} \Delta_{-}\right\rangle$ are connected to the correlators $a_{++}\left(a_{+-}\right)$ by:
\begin{equation}\label{equ.13}
a_{++}=a_{--}=\frac{1}{N_{+}^{2}} \frac{\pi^{2}}{16}\left\langle\Delta_{ \pm}^{2}\right\rangle,
\end{equation}
\begin{equation}\label{equ.14}
a_{+-}=a_{-+}=\frac{1}{N_{+} N_{-}} \frac{\pi^{2}}{16}\left\langle\Delta_{+} \Delta_{-}\right\rangle,
\end{equation}
where $N_{\pm}$ is the total number of positively or negatively charged particles in the corresponding $\eta$ interval.

Early studies of charge separation fluctuations perpendicular to the reaction plane in high energy physics experiments used the three-point correlator $\gamma \equiv  \left\langle\left\langle\cos \left(\phi_{\alpha}+\phi_{\beta}-2\Psi_{\mathrm{RP}}\right)\right\rangle\right\rangle$, where the double averaging is done over all particles in an event and over all events \cite{Abelev:2009ac,Abelev:2009ad,Abelev:2012pa}. Unfortunately, the $\gamma$ correlator includes some background contributions not related to CME \cite{Bzdak:2010fd,Bzdak:2012ia,Liao:2015Pr}. The background contribution is mainly from the elliptic flow $\left(v_{2}\right)$ in combination with the two-particle correlations. The two-particle correlator $\delta \equiv\left\langle\cos \left(\phi_{\alpha}-\phi_{\beta}\right)\right\rangle$ was introduced to solve this problem.

By inducing $H$ and $F$ as CME and no CME background contribution, one can express  $\gamma$ and  $\delta$ in the following way \cite{Bzdak:2012ia,Liao:2015Pr}.
\begin{equation}\label{equ.15}
\gamma \equiv\left\langle\cos \left(\phi_{1}+\phi_{2}-2 \Psi_{R P}\right)\right\rangle= k v_{2} F-H
\end{equation}
\begin{equation}\label{equ.16}
\delta \equiv\left\langle\cos \left(\boldsymbol{\phi}_{1}-\boldsymbol{\phi}_{2}\right)\right\rangle= F+H
\end{equation}
The $H$ factor related to chiral magnetic signal can be obtained as:
\begin{equation}\label{equ.17}
H^{k}=\left(k v_{2} \delta-\gamma\right) /\left(1+k v_{2}\right)
\end{equation}

where coefficient $\kappa$ ranges from $1$ to $2$, due to the finite detector acceptance and theoretical uncertainties \cite{Bzdak:2012ia,Liao:2015Pr}; we take the experimental results with $\kappa = 1.5$ in the following. A one-to-one correspondence is made between the charge separations $a_{++}\left(a_{+-}\right)$ of the KWM model and the experimental results $H_{\mathrm{SS}}\left(H_{\mathrm{OS}}\right)$. Therefore, the calculated result $a_{++}-a_{+-}$ can be compared with the experimental result $H_{\mathrm{SS}}-H_{\mathrm{OS}}$, as shown in Fig.~\ref{fig.3}.

\end{multicols}
\begin{center}
\includegraphics[width=18cm]{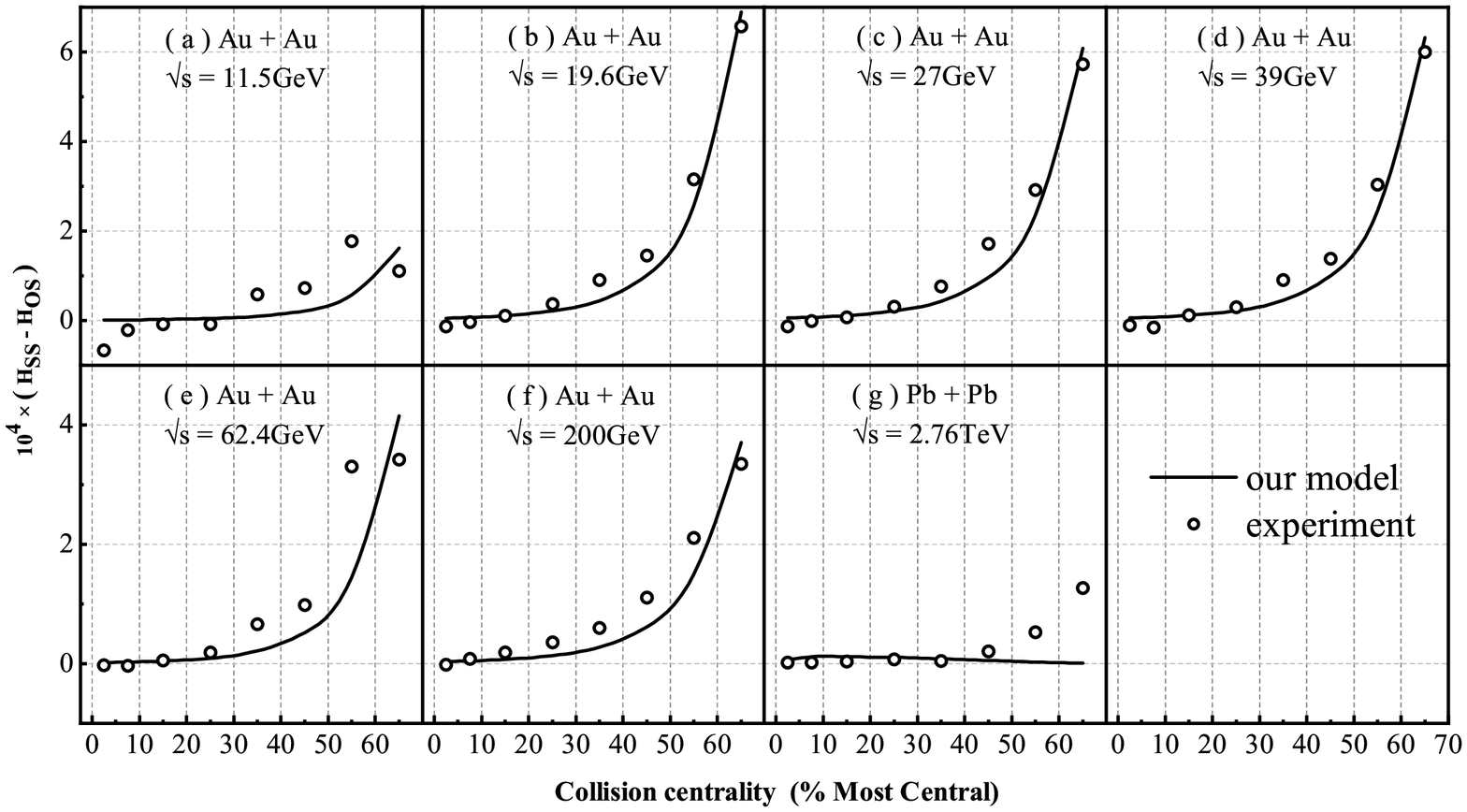}
\figcaption{\label{fig.3} Comparison between the centrality dependence of $a_{++}-a_{+-}$ from our model and that of the background subtracted experimental observable $H_{\mathrm{SS}}-H_{\mathrm{OS}}$ at the RHIC and LHC energies\cite{Adamczyk:2014mzf}.}
\end{center}
\begin{multicols}{2}

It can be seen from Fig.~\ref{fig.3} that the model explains better the experimental data at the energies of RHIC BES and the top RHIC energy than at the LHC energy. For the Au-Au collisions at RHIC, the CME signal given by our calculations increases from the central to peripheral collisions, and the general trend of our results is consistent with the experiment. However, for the Pb-Pb collisions at the LHC energy of $\sqrt{s}=2760$ GeV, the experimental CME signal~\cite{Abelev:2012pa} is very small, and only a small signal is present at the centrality of $60\%-70\%$. Our model predicts no CME signal in this case.

Fig.~\ref{fig.4} shows $H_{\mathrm{SS}}-H_{\mathrm{OS}}$ as a function of beam energy for two centrality bins at the RHIC energies. The experimental results with $\kappa = 1.5$  from Ref. \cite{Adamczyk:2014mzf} are used as reference for our theoretical calculations.
The results in Fig.~\ref{fig.4} show that our calculated CME signal has a very similar trend  as the  experimental measurements. The magnitude of our predictions is lower than the experimental data, presumably because our magnetic field decreases very quickly, as shown in Fig.~\ref{fig.2}. A quickly decreasing trend in the interval from 19.6 GeV to 7.7 GeV is seen, which suggests that hadronic interactions dominate over partonic interactions at low beam energies. Generally speaking, our model closely follows the evolution of the magnetic field, so the results of our calculations include certain model limitations.

\begin{center}
	\includegraphics[width=8cm]{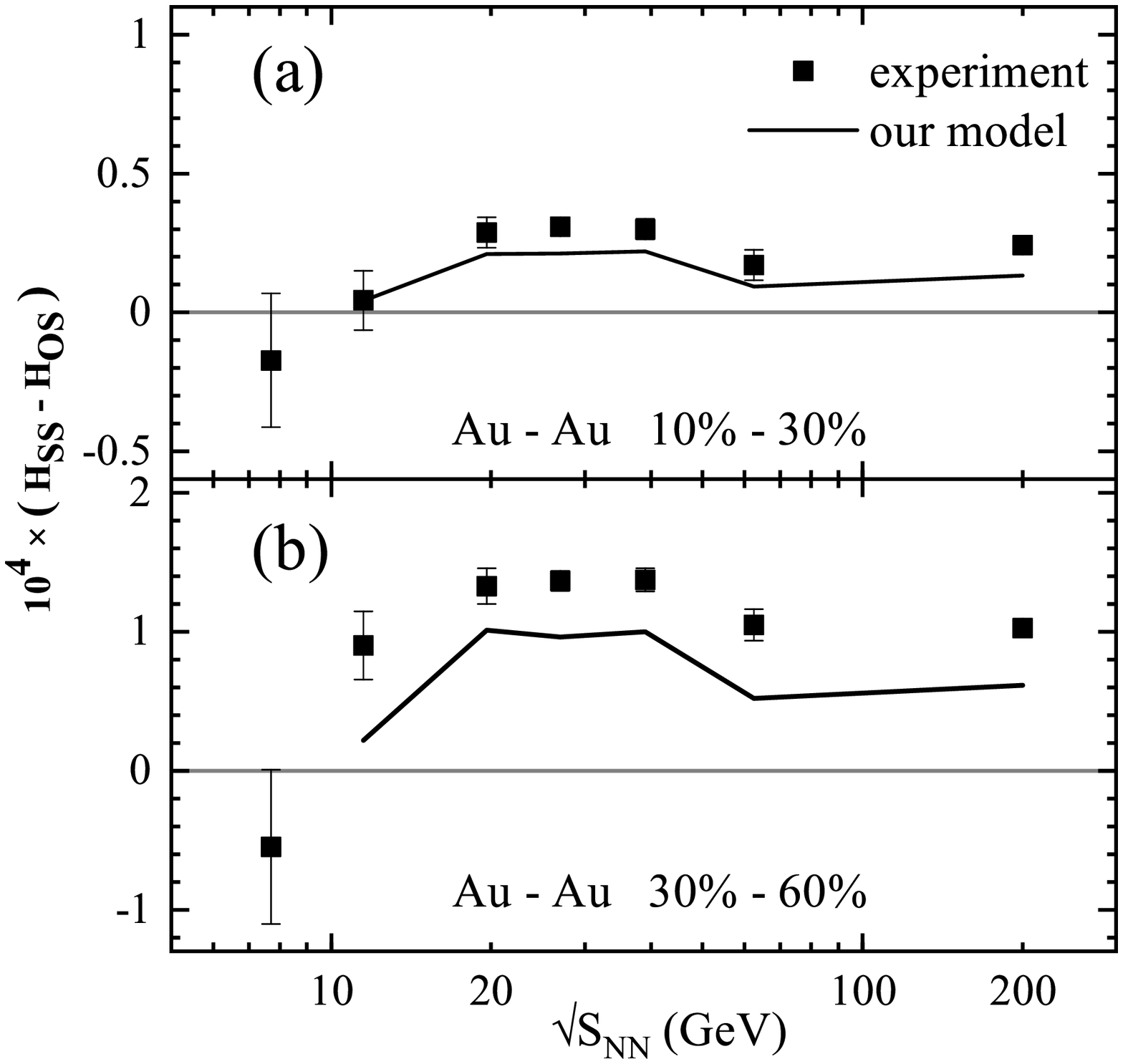}
	\figcaption{\label{fig.4} $H_{\mathrm{SS}}-H_{\mathrm{OS}}$ as a function of beam energy for two
		centrality bins at the RHIC energies. The solid
		curves are our calculation results. The experimental
		results are from Ref.\cite{Adamczyk:2014mzf} with $\kappa = 1.5$. }
\end{center}

\section{Chiral magnetic current}\label{cmc}
Let us now turn to the induced chiral magnetic current generated by the magnetic field in relativistic heavy-ion collisions at the RHIC and LHC energies. Assuming that the generated magnetic field has a homogeneous distribution, one can calculate the induced current as \cite{She:2017icp,Kharzeev:2009pj}:
\begin{equation}\label{equ.18}
j(t)=\int_{0}^{\infty} \frac{\mathrm{d} \nu}{\pi} \tilde{B}(\nu) \left[\sigma_{\chi}^{\prime}(\nu) \cos (\nu t)+\sigma_{\chi}^{\prime \prime}(\nu) \sin (\nu t)\right] ,
\end{equation}
where $\nu$ is the frequency, and the Fourier transform of the magnetic field is given by
\begin{equation}\label{equ.19}
\tilde{B}(\nu)=\int_{t_{0}}^{\infty} \mathrm{d} t B(t) e^{i \nu t}.
\end{equation}

The real $\sigma_{\chi}^{\prime}(\nu)$ and imaginary $\sigma_{\chi}^{\prime \prime}(\nu)$ parts of the chiral magnetic conductivity are related by the Kramers-Kroning relation
\begin{equation}\label{equ.20}
\sigma_{\chi}^{\prime}(\nu)=\frac{1}{\pi} \mathcal{S} \int_{-\infty}^{\infty} d q_{0} \frac{\sigma_{\chi}^{\prime \prime}\left(q_{0}\right)}{q_{0}-\nu},
\end{equation}
\begin{equation}\label{equ.21}
\sigma_{\chi}^{\prime \prime}(\nu)=-\frac{1}{\pi} \mathcal{S} \int_{-\infty}^{\infty} d q_{0} \frac{\sigma_{\chi}^{\prime}\left(q_{0}\right)}{q_{0}-\nu},
\end{equation}
where $\sigma_{\chi}(\nu)=\lim _{\vec{p} \rightarrow 0} \sigma_{\chi}\left(p_{0}=\nu, \vec{p}\right)$. The symbol $\mathcal{S}$ in eqs.~(\ref{equ.20}) and (\ref{equ.21}) defines the integral of a singularity in the upper and the lower complex plane. The chiral magnetic conductivity should be complex, and is given as
\begin{equation}\label{equ.22}
\sigma_{\chi}(p)=\sigma_{\chi}^{\prime}(p)+i \sigma_{\chi}^{\prime \prime}(p),
\end{equation}
where both $\sigma_{\chi}^{\prime}(p)$ and $\sigma_{\chi}^{\prime \prime}(p)$ are real functions. They can be expressed as:
\begin{equation}\label{equ.23}
\sigma_{\chi}^{\prime}(p)=\frac{1}{p^{i}} \operatorname{Im} R_{R}^{i}(p),
\end{equation}
\begin{equation}\label{equ.24}
\sigma_{\chi}^{\prime \prime}(p)=-\frac{1}{p^{i}} \operatorname{Re} R_{R}^{i}(p),
\end{equation}
where $R_{R}^{i}(p)=\frac{1}{2} \varepsilon^{i j k} \widetilde{\Pi}_{R}^{j k}(p)$ is the retarded correlator, which can be calculated as
\begin{align}\label{equ.25}
\begin{aligned} R_{R}^{i}(p)=& \frac{i e^{2}}{16 \pi^{2}} \frac{p^{i}}{p} \frac{p^{2}-p_{0}^{2}}{p^{2}} \int_{0}^{\infty} \mathrm{d} q g(q)\sum_{t=\pm}\left(2 q+t p_{0}\right)
\\ & \times \log\left[ \frac{\left(p_{0}+i \varepsilon+t q\right)^{2}-(q+p)^{2}}{\left(p_{0}+i \varepsilon+t q\right)^{2}-(q-p)^{2}}\right],
\end{aligned}
\end{align}
where
\begin{equation}\label{equ.26}
g(q)=\sum_{s=\pm} s\left(\tilde{n}\left(q-\mu_{s}\right)-\tilde{n}\left(q+\mu_{s}\right)\right),
\end{equation}
and $\tilde{n}(x)=[1+\exp (\beta x)]^{-1}$ is the Fermi-Dirac distribution function. One can compute the imaginary part of the logarithm in Eq.~(\ref{equ.25}) with $\mathrm{p}=|\vec{p}| \geq 0$ and $\mathrm{q} \geq 0$ as
\begin{equation}\label{equ.27}
\begin{split}
&\operatorname{Im} \sum_{t=\pm}\left(2 q+t p_{0}\right) \log \frac{\left(p_{0}+i \varepsilon+t q\right)^{2}-(q+p)^{2}}{\left(p_{0}+i \varepsilon+t q\right)^{2}-(q-p)^{2}} =\\  & \pi\left[2 q-\left|p_{0}\right| \theta\left(p_{0}^{2}-p^{2}\right)\right]\left[\theta\left(q_{+}-q\right)-\theta\left(q_{-}-q\right)\right]\\  & +\pi p_{0} \theta\left(p^{2}-p_{0}^{2}\right)\left[\theta\left(q-q_{+}\right)-\theta\left(q-q_{-}\right)\right],
\end{split}
\end{equation}
where $q_{ \pm}=\frac{1}{2}\left|p_{0} \pm p\right|$.

After computing the real and imaginary parts of the magnetic conductivity, we use Eq.~(\ref{equ.18}) to calculate the electromagnetic current. In order to use Eqs.~(\ref{equ.18}) and (\ref{equ.19}), we need the dependence of the magnetic field on time after the formation of a parton. The magnetic field evolution for $t \geq t_{0}$ is given by Eq.~(\ref{equ.01}).
\end{multicols}
\begin{center}
\includegraphics[width=18cm]{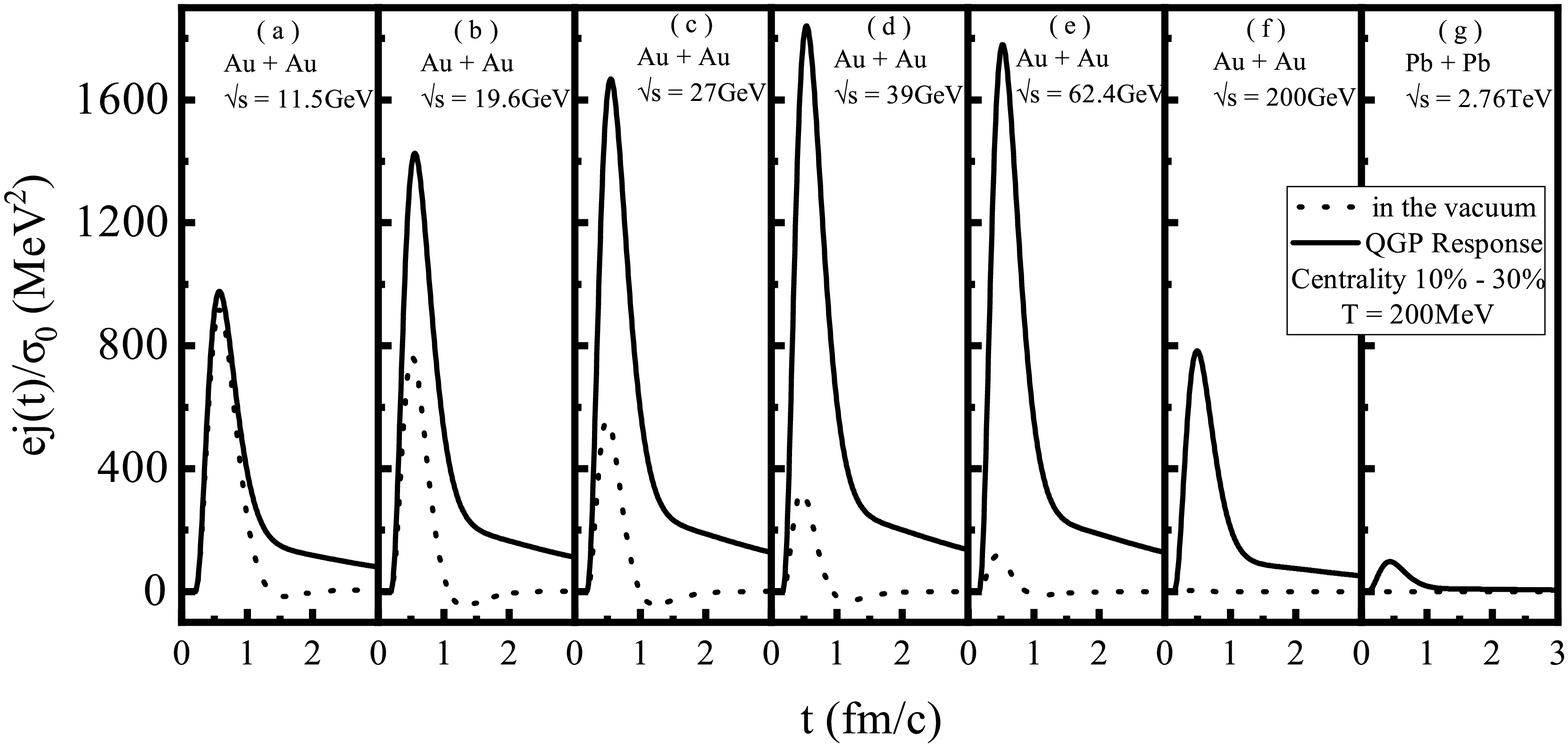}
\figcaption{\label{fig.5} Time evolution of the induced electromagnetic current, normalized to zero frequency chiral magnetic conductivity $(\sigma_{0})$, fot the RHIC and LHC collision energies. The solid curves show the results with the QGP response, and the dashed curves in vacuum. The centrality is $10\% \sim 30 \%$.}
\end{center}
\begin{multicols}{2}
\end{multicols}
\begin{center}
\includegraphics[width=18cm]{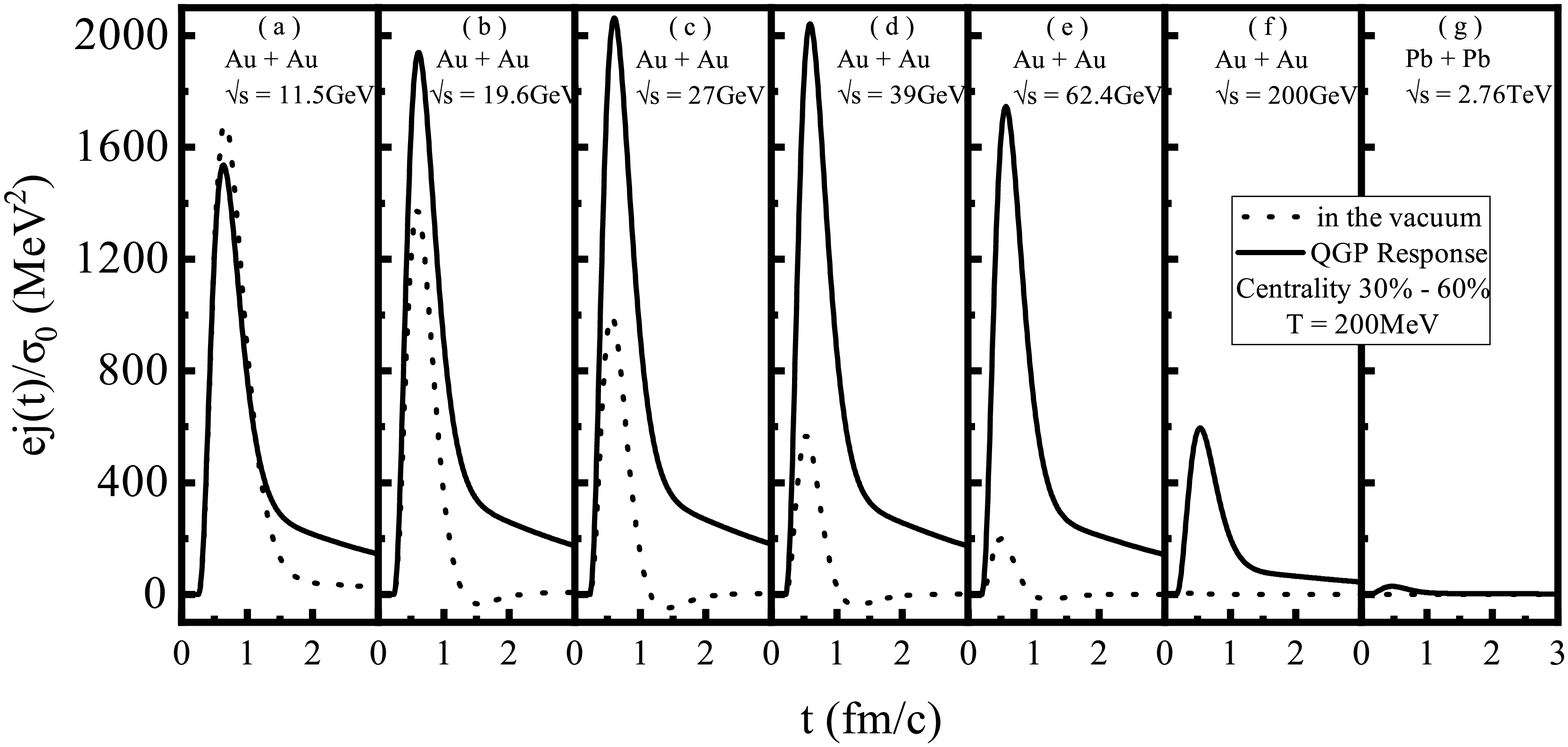}
\figcaption{\label{fig.6} As Fig.5, but for the centrality of $30\% \sim 60\%$.}
\end{center}

\begin{multicols}{2}
Fig.~\ref{fig.5} shows the time dependence of the induced electromagnetic current, normalized to the zero frequency chiral magnetic conductivity $\left(\sigma_{0}\equiv\sigma_{\chi}(\omega=0)=\frac{e^{2}}{2 \pi^{2}} \mu_{5}\right)$, for collisions with the centrality of $10\% \sim 30 \%$ at the RHIC and LHC energies.
It can be observed that the electromagnetic current signal manifests as a strong pulse, which reaches a maximum at $t\sim$ 1fm. The maximum value of the electromagnetic current signal directly reflects the intensity of the induced electromagnetic current. It increases with the collision energy from $\sqrt{s} = 19.6 \textrm{GeV}$ to $\sqrt{s} = 39 \textrm{GeV}$,  remains almost unchanged from $\sqrt{s} = 39 \textrm{GeV}$ to $\sqrt{s} = 62.4 \textrm{GeV}$, and then decreases from $\sqrt{s} = 39 \textrm{GeV}$ to $\sqrt{s} = 2760 \textrm{GeV}$.

Fig.~\ref{fig.6} is the same as Fig.~\ref{fig.5}, but for the centrality of $30\% \sim 60\%$. It can be seen that the maximum value of the electromagnetic current increases from $\sqrt{s} = 19.6 \textrm{GeV}$ to $\sqrt{s} = 39 \textrm{GeV}$, and then decreases from $\sqrt{s} = 39 \textrm{GeV}$ to $\sqrt{s} = 2760 \textrm{GeV}$. Figures.~\ref{fig.5} and \ref{fig.6} both indicate that the intensity of the induced electromagnetic current is clearly larger with the QGP response than in vacuum, and both show that the CME signal almost vanishes at the LHC energy $\sqrt{s} = 2760 \textrm{GeV}$.

The dependence of the time-integrated current signal $\left(\mathrm{Q}=\int \mathrm{j}(\mathrm{t}) \mathrm{dt}\right)$ on the center-of-mass energy at RHIC and LHC is shown for two centralities in Fig.~\ref{fig.7}(a, b). It is found that the time-integrated current signal reaches a maximum around $\sqrt{s} \approx 39$ GeV, and then decreases with $\sqrt{s}$. The qualitative trends of Figs.~\ref{fig.5}, \ref{fig.6} and \ref{fig.7} are in agreement with the CME experimental results obtained at RHIC and  LHC in a wide range of beam energies \cite{Adamczyk:2014mzf}.

\begin{center}
	\includegraphics[width=8cm]{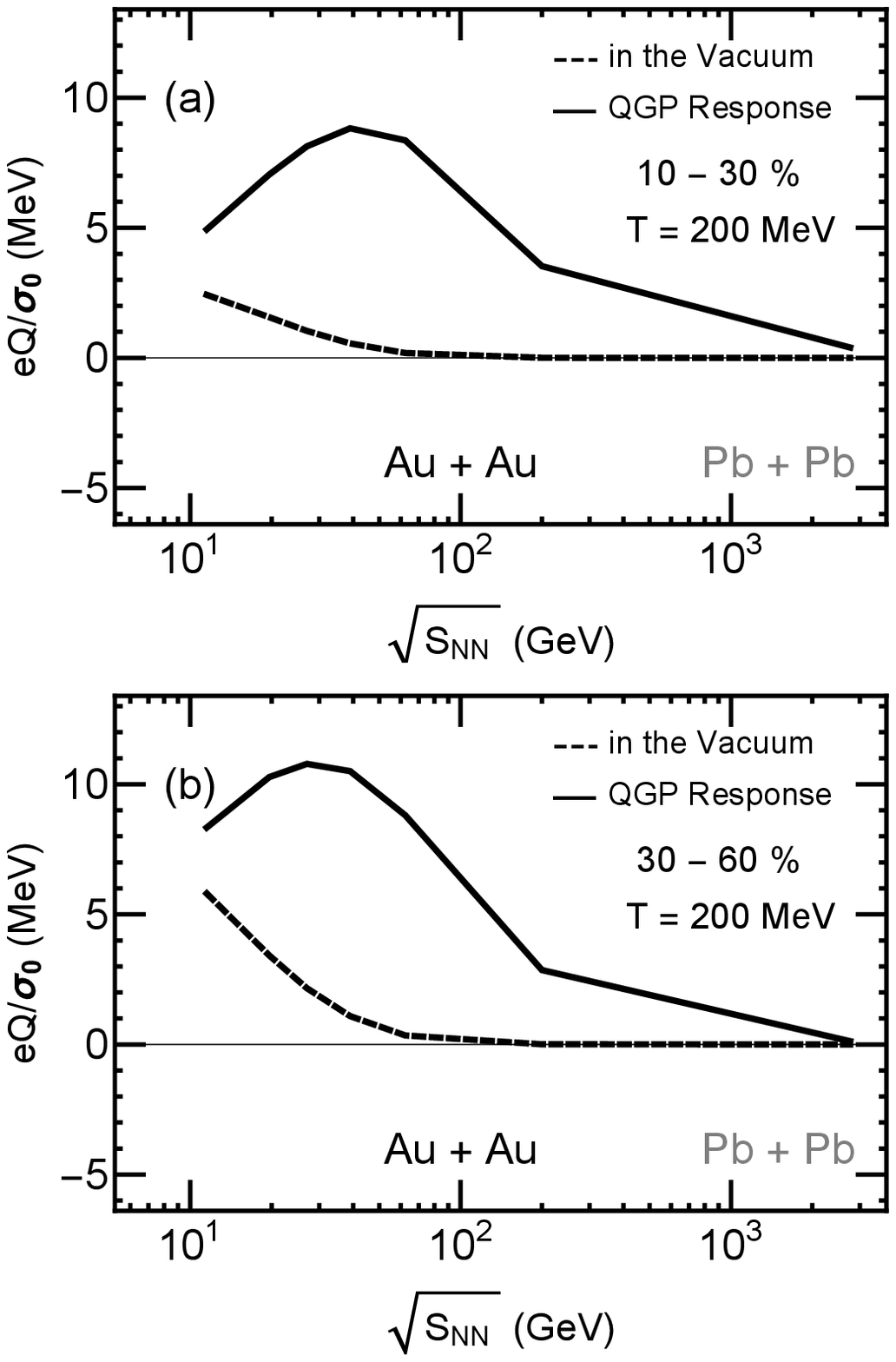}
	\figcaption{\label{fig.7} Dependence of the time-integrated current signal $\left(\mathrm{Q}=\int \mathrm{j}(\mathrm{t}) \mathrm{dt}\right)$ on collision energy at RHIC and LHC for the centrality of $10\%  \sim  30 \%$ (a), and the centrality $30\% \sim 60\%$ (b). The solid curves are the results with the QGP response  and the dashed lines in vacuum. }
\end{center}
\section{Summary}\label{Summary}
Considering the magnetic field response of the QGP medium, we performed a systematical study of the charge separation  and compared it with the experimental results for the background-subtracted correlator $H$ at the RHIC and LHC energies. The results show that our calculated chiral magnetic effect signal has same trend as the experimental results at RHIC. Quantitatively, our results from appear to be lower than the experimentally measured correlations, which may be due to the rapid decrease of the magnetic field.

The time evolution of the chiral electromagnetic current at the energies of the RHIC Beam Energy Scan and the LHC energy was systematically investigated. The dependence of the time-integrated current signal on the center-of-mass energy $\sqrt{s}$ at RHIC and LHC and different centralities was also studied.
 In such a wide range of collision energies, it is important to identify the collision energy at which the electromagnetic current is largest, so as to help steer the experimental study of CME. Our phenomenological analysis showed that the time-integrated electromagnetic current has a maximum $\sqrt{s} \approx 39$ GeV
 The qualitative trend of the induced electromagnetic current with collision energy is in agreement with the CME experimental results from RHIC and LHC \cite{Adamczyk:2014mzf}. We argue that the electromagnetic current at the LHC energy $\sqrt{s} = 2760 \textrm{GeV}$ is so small that CME cannot be produced.

\section*{Acknowledgments}
Supported by National Natural Science Foundation of China (11875178, 11475068, 11747115),the CCNU-QLPL Innovation Fund (QLPL2016P01) and the Excellent Youth Foundation of Hubei Scientific Committee (2006ABB036)
\end{multicols}

\begin{multicols}{2}

\end{multicols}
	
\clearpage

\end{CJK*}

\end{document}